\documentstyle[epsf,prd,aps,psfig,floats]{revtex}

\newcommand{\be}{\begin{equation}}\newcommand{\ee}{\end{equation}}
\newcommand{\bea}{\begin{eqnarray}}\newcommand{\eea}{\end{eqnarray}}
\newcommand{\brr}{\begin{array}}\newcommand{\err}{\end{array}}
\newcommand{\bit}{\begin{itemize}}\newcommand{\eit}{\end{itemize}}
\newcommand{\ben}{\begin{enumerate}}\newcommand{\een}{\end{enumerate}}

\newcommand{\ba}{\begin{array}}
\newcommand{\ea}{\end{array}}

\def\lrar{\leftrightarrow}

\def\non{\nonumber}\def\ran{\rangle}

\def\te{\theta}

\def\om{\omega}

\def\1{{_{1}}}\def\2{{_{2}}}
\def\bk{{\bf {k}}}

\begin{document}
\twocolumn[\hsize\textwidth\columnwidth\hsize\csname@twocolumnfalse\endcsname

\title{Neutrino mixing and Lorentz invariance}

\author{Massimo Blasone$^{\sharp}$, Jo\~ao Magueijo$^{\flat}$ and
Paulo Pires-Pacheco$^{\flat}$\vspace{3mm}}

\address{${}^{\sharp}$
Institute f\"ur Theoretische Physik, Freie Universit\"at Berlin,
Arnimallee 14, D-14195 Berlin, Germany
\\ ${}^{\flat}$
Blackett  Laboratory, Imperial  College London,   Prince Consort Road,
London  SW7 2BW, U.K. }

\maketitle

\begin{abstract}
We use previous work on the Hilbert space for mixed fields
to derive deformed dispersion relations for neutrino
flavor states. We then discuss how these dispersion relations may
be incorporated into frameworks encoding the breakdown of Lorentz
invariance. We consider non-linear relativity schemes (of which
doubly special relativity is an example), and also frameworks
allowing for the existence of a preferred frame. In both cases we
derive expressions for the spectrum and end-point of beta decay,
which may be used as an experimental probe of the peculiar way in which
neutrinos experience Lorentz invariance.
\end{abstract}
\pacs{PACS Numbers: *** }

]
\renewcommand{\thefootnote}{\arabic{footnote}}
\setcounter{footnote}{0}

\vspace{8mm}

%\section{Introduction}
The subject of neutrino oscillations has now matured from an
insightful prediction by Bruno Pontecorvo~\cite{Pont} and the
early results of Homestake~\cite{homestake} to a structured
framework backed by a wealth of new quantitative  data
\cite{Kam,SNO,kamland,K2K}. These advances have been paralleled by
much progress on the theoretical front, both in the phenomenological
pursuit of more refined oscillation formulae and in efforts
to give the theory a sound formal structure
within Quantum Field Theory (QFT).

A major outstanding  question  is that of the existence of a
Hilbert space for the flavor states~\cite{kimbook}. Pontecorvo's
treatment of flavor states as Quantum Mechanical superpositions is
forbidden by the Bargmann super-selection rules~\cite{Barg} (see, however,
Ref.\cite{greenberger}). This
problem only found its resolution with a full QFT
treatment~\cite{BV95,BHV99,hannabuss,Ji,fujii1,bosonmix}. In
the resulting picture, flavor states -- rather than mass
eigenstates -- constitute the real physical entities.
On this basis oscillation formulae were derived which exhibit
corrections with respect to the usual
ones \cite{BHV99,bosonmix,Ji,fujii1}.
%This was
%further supported by the discovery of an associated Berry
%phase~\cite{Berry}, an extension to three-flavors \cite{3flav} and
%bosons~\cite{bosonmix}, and the construction of the corresponding
%four-currents\cite{currents}.

In this letter we elaborate on the curious fact that neutrino
flavor states don't satisfy the standard dispersion relations
$E^2-k^2=m^2$, where $E$ is energy, $k$ momentum and $m$ rest
mass. This follows trivially from the fact that flavor states are
superpositions of mass eigenstates, which do satisfy standard
dispersion relations, but with different masses.

Such a peculiarity leads to a formal connection between neutrino
oscillations  and a very different field. Deformed dispersion
relations have been used as a phenomenological framework for
quantum gravity~\cite{amel,amel1,liouv}, capable of explaining
high-energy cosmic ray
anomalies~\cite{review,crexp,cosmicray,leejoao1}, and of
establishing an observer independent border between the classical
and quantum pictures of space-time~\cite{amelstat,gli,leejoao}.

Thus it is possible to employ a formalism describing violations of
Lorentz invariance in quantum gravity to examine how neutrino
flavor states must experience some form of breakdown of standard
Lorentz invariance. Deformed dispersion relations signal either
the presence of a preferred frame~\cite{amel,amel1} or a
non-linear realization of Lorentz invariance~\cite{leejoao} (for
simplicity we exclude the possibility of quantum groups). We
consider how neutrino flavor states might fit both possibilities,
focusing mainly on the second.

We restrict ourselves to the simplest case,
i.e. two-flavor mixing in the
Pontecorvo approximation, and leave
the full QFT treatment and the extension to three flavors to a longer
publication~\cite{bmp}. We identify the non-linear Lorentz
transformations which leave the neutrino deformed dispersion
relations frame-independent. We then use these results to work out
energy conservation formulae, with and without a preferred frame.
These lead to distinct predictions for the spectrum of beta decay,
which may thus be used as a test of Lorentz invariance. We also
discuss the meaning of a possible negative
mass squared, as suggested by some observations~\cite{troi,mainz}.

Consider the mixing relations for two flavors:
\bea
\non \nu_e(x) & = &\cos\theta \,\,\nu_1(x) +
\sin\theta\,\, \nu_2(x)
\\[2mm] \label{fermix}
\nu_{\mu}(x) & =& -
\sin\theta\,\, \nu_1(x) + \cos\theta\,\,\nu_2(x)
\eea
Without loss of generality we take $0\le\te\le\frac{\pi}{4}$ and
$m_2 > m_1$. As explained in~\cite{BV95}, it is possible to define
a vacuum state, creation and annihilation operators, and a Hilbert
space for flavor states. The properly defined flavor states
\cite{BHV99,bosonmix} are then eigenstates  of the
flavor charge and of the momentum operators \footnote{For example,
the electron neutrino state is defined by
$Q_e |\nu_e\ran =  |\nu_e\ran$
and ${\bf P}_e |\nu_e\ran =  {\bf k} |\nu_e\ran $, with
$Q_e=\int d^3x \nu_e^\dag(x) \nu_e(x)$ and ${\bf P}_e =
\int d^3{\bf x} \nu^\dag_e(x) (-i\nabla)\nu_e(x)$.}.
Obviously they are
not eigenstates of the Hamiltonian, however it makes sense to
consider the expectation value of $H$ on the flavor states and
define from it dispersion relations. In this paper we consider the
simplest case, i.e.  we limit ourselves to the usual Pontecorvo
states, which are a good approximation for the full QFT flavor
states when the masses are sufficiently close to each other.
%\footnote{The QFT flavor states treatment~\cite{bmp} adds an
%extra term to the Pontecorvo states (\ref{fermix}), but it does
%not affect our argument and so we will ignore it here.}.
We then have
\bea\non |\nu_e\ran& = &\cos\theta \,\,|\nu_1\ran +
\sin\theta\,\, |\nu_2\ran
\\ [2mm] \label{Pontec}
|\nu_{\mu}\ran & =& -
\sin\theta\,\, |\nu_1\ran + \cos\theta\,\,|\nu_2\ran   \eea
If we compute the expectation value of the Hamiltonian on the
flavor states (\ref{Pontec}) we find the result:
\bea\non E_e &\equiv & \langle\nu_e| H|\nu_e \rangle  = \om_{k,1}
\cos^2\te + \om_{k,2} \sin^2 \te
\\ [2mm] \label{dispersion}
E_\mu &\equiv& \langle\nu_\mu| H|\nu_\mu \rangle  =  \om_{k,2}
\cos^2\te + \om_{k,1} \sin^2 \te
\eea
where $H|\nu_i\rangle=\omega_i|\nu_i\rangle$ ($i=1,2$) and \be
\omega_{k,i} = \sqrt{ {\mathbf k}^2+m_i^2}\,  .\ee Since
the sum of two square roots is generally not a square root,
we find that, except for trivial cases, flavor states do not satisfy the
usual dispersion relations. Obviously the energies in
Eq.~(\ref{dispersion}) are only expectation values subject to
fluctuations but it is nevertheless sensible to consider the
modified Lorentz transformation for the classical limit of the
theory.

It is immediately obvious that the minimal energy of a flavor
state is achieved at zero momentum and is
\bea m_e &\equiv&
E_e(\bk =0)\,=\, m_1 \cos^2\te + m_2 \sin^2 \te\label{restmass}
\\ [1.5mm]
m_\mu &\equiv& E_\mu(\bk =0)\,=\,
m_2 \cos^2\te + m_1 \sin^2 \te \label{restmass1}
\eea
However this ``mass'' is not the kinematic invariant, since the
dispersion relations Eq.~(\ref{dispersion}) are no longer
invariant under the usual Lorentz transformations. Instead, if we
are to avoid introducing a preferred frame, neutrinos must feel a
non-linear representation of the Lorentz group as constructed
by~\cite{leejoao,leejoao1}. The dispersion relations
Eq.~(\ref{dispersion}) may be put in the form used in that work:
\bea \label{lorentz} E_e^2 \, f_e^2(E_e)\, - \,\bk^2
\,g_e^2(E_e)\, =\,  M_e^2
\\ [2mm] \label{lorentz2}
E_\mu^2 \, f_\mu^2(E_\mu)\, -\, \bk^2\, g_\mu^2(E_\mu)\, =\,
M_\mu^2
\eea
from which the recipes given in~\cite{leejoao1} are
straightforward to apply. Eq.~(\ref{dispersion}) leads to
\bea
\label{eqe1}
 \left(E_e^2 - \om_1^2 \cos^4\te - \om_2^2 \sin^4 \te\right)^2 =  4 \om_1^2
 \om_2^2 \sin^4 \te \cos^4\te
 \\ [1.5mm]
\left(E_\mu^2 - \om_2^2 \cos^4\te - \om_1^2 \sin^4 \te\right)^2 =
4 \om_1^2
 \om_2^2 \sin^4 \te \cos^4\te
 \eea
and, for example, Eq.\,(\ref{eqe1}) can be rewritten as a quadratic
equation in $\bk^2$, of the form $A\bk^4 + B\bk^2 +C=0$, with
 \bea A&=& \cos^2 2\te \label{A}\\
 B&=&2\cos (2\te) {\tilde m}^2_-
-E_e^2\left(1 + \cos^2 2\te\right) \\
C &=&E_e^4 - 2 E_e^2 {\tilde m}^2_+ +{\tilde m}^4_-
\eea
and ${\tilde m}^2_\pm= m_1^2 \cos^4\te \pm m_2^2 \sin^4 \te$.
Assuming first that $\theta\ne\pi/4$, its solutions are
\bea \label{solution}
\bk^2_{\pm}&=&{E_e^2\left(1 + \cos^2 2\te\right)-2{\tilde m}^2_-\cos 2\te
\pm {\sqrt\Delta}\over 2\cos^2 2\te}\\[1.5mm]
\Delta&=&E_e^2\sin^4 2\te\left(E_e^2+(m_2^2-m_1^2)\cos 2\te\right)
\eea
Since $E$ is a monotonically growing function of $k$ with minimum
$m_e$,  the discriminant satisfies $\Delta>0$,
 so that the roots are guaranteed to be real. The physically
relevant solution is $\bk^2_{-}$ (e.g. study the special case $ m
= m_1 = m_2$).
Comparing (\ref{solution}) and (\ref{lorentz})
leads to:
\bea 2f^2_e(E_e) &=&1+ \frac{1} {\cos^2(2\te) }
\nonumber
\\
&&- \frac { \sqrt{E_e^2 + \left(m_2^2 - m_1^2\right) \cos(2\te)
}} {E_e} \tan^2(2\te)
\\
g^2_e(E_e)& =& 1
\\
M_e^2&=&\frac {{\tilde m}_-^2}{\cos 2\te}=
\frac{m_1^2\cos^4\te -m_2^2\sin^4\te} {\cos (2\te)}
\eea
Notice that $M_e^2< 0$ for
\be \label{tachcond}
 \tan\theta> \sqrt{m_1/m_2}
\ee
This does not imply causality
violations, since it can be shown (directly from
Eq.~(\ref{dispersion})) that the neutrino velocity (defined either
as $v=p/E$ or as $v=dE/dp$) is a monotonically growing function of
its momentum, with $v<1$. However other aspects of tachyonic
behavior will be identified later.

The case of maximal mixing $\theta=\pi/4$ has to be treated separately.
Then $A=0$ in Eq.~(\ref{A}), collapsing the order of the equation.
One then finds trivially
\bea f^2_e(E_e) &=&1+ \left(\frac{m_1^2-m_2^2} {4E_e^2}\right)^2\\
g^2_e(E_e)& =& 1\\
M_e^2&=&\frac {m_1^2+m_2^2} {2}
 \eea
Similar expressions, but with $m_1 \lrar m_2$, apply to muon neutrinos.
In the whole parameter space $M_\mu^2>0$.

It is now possible to identify the non-linear realization of the
Lorentz group which leaves these dispersion relations invariant.
They are generated by the transformation $ U \circ (E, {\bf
k})=(Ef,{\bf k}g) \label{udef} $ applied to the standard
Lorentz generators ($L_{ab} = p_a {\partial \over \partial p^b} -
 p_b {\partial \over \partial p^a}$):
\begin{equation}\label{U}
K^i = U^{-1} [p_0] L_0^{\ i} U [p_0 ]\, .
\end{equation}
This amounts to requiring linearity for the auxiliary variables ${\tilde
E}=Ef(E)$ and ${\tilde k}=kg(E)$. The resulting non-linear transformations
for $E$ and $k$ are a non-linear representation of the Lorentz group
ensuring that the deformed dispersion relations found for
flavor states are valid in all frames.
For $\nu_e$ the transformation (\ref{U}) only exists for
\be\label{emin}
E_e^2\ge E_{min}^2=
(m_2^2-m_1^2)\frac{\sin^4 2\theta}{4\cos 2\theta}
\ee
However, if $M_e^2\ge 0$, one has $E_{min}<E_e(k=0)=m_e$, so that
the transformation exists for all allowed values of $E_e$. This is
not the case if $M_e^2<0$. Then one must impose
$|k|>k_{min}=\sqrt{-M_e^2}$ to have a well defined transformation
(\ref{U}). Since this condition is frame-independent, no conflict
with the principle of relativity arises (notice also
that $E\ge 0$ always transforms into $E\ge 0$).

The ugly alternative to non-linear realizations is that neutrino
flavor states are subject to the usual linear Lorentz
transformations, so that their deformed dispersion relations
select a preferred frame (typically the cosmological
frame). With or without
the introduction of a preferred frame, flavor states are at odds
with conventional Lorentz invariance.

We now explore the physical implications of these peculiarities,
taking as an example beta decay:
$A\rightarrow B + e^- +{\bar \nu}_e$
where $A$ and $B$ are two nuclei (e.g. $^3$H and $^3$He).
Given that the nuclei and the electron satisfy linear Lorentz
transformations, and that $E_{e}f_e(E_{e})$
transforms linearly (cf.
Eq.~(\ref{U})), the only covariant law of energy conservation
is
\be E_A=E_B+E +E_{e}f_e(E_{e})\, .
\label{ennonli}
\ee
where $E$ is the electron energy, and as before the subscript $e$
refers to the electron neutrino.
If, on the contrary, we insist upon the standard
law \be E_A=E_B+E+E_{e}\label{enpref}\ee we have
introduced a preferred frame, and are in conflict with the principle
of relativity.

These two choices are reflected in different predictions for the
endpoint of $\beta$ decay, that is, the maximal kinetic energy
($K_{max}$) the electron can carry away. This is constrained by
the available energy $Q=E_A-E_B-m\approx m_A-m_B-m$, where $m$ is the electron mass.  For the
tritium decay, $Q=18.6$ KeV. $Q$ is shared between the
(unmeasured) neutrino energy and the (measured) electron kinetic
energy $K$. If the neutrino were massless, then $K_{max}=Q$. If
the neutrino were a mass eigenstate (say with $m_e=m_1$), then
$K_{max}=Q-m_1$.

\begin{figure}[tbp]
\psfig{file=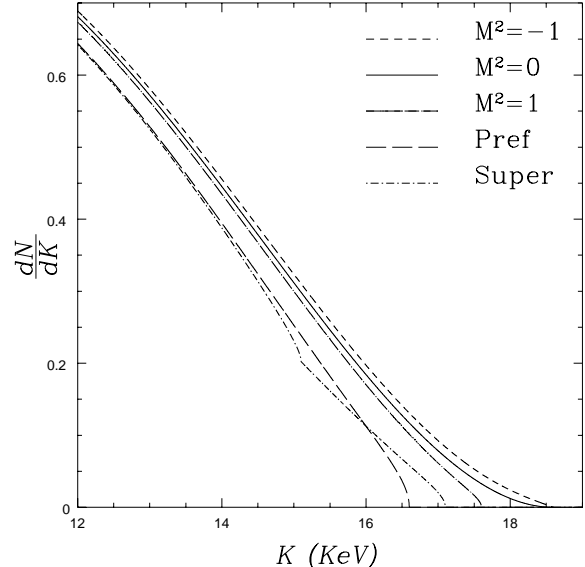,width=8cm} \caption{The tail of the tritium
$\beta$ spectrum for a massless neutrino (solid) and for Lorentz
invariant flavor states with $M_e^2=\pm 1$ KeV$^2$ (dashed and
dot-long dashed lines). We have also plotted the case of neutrinos
with a preferred frame and $m_1=1.5$ KeV, $m_2=3.5$ KeV, $\theta=30^\circ$
(for which $M_e=1$ KeV but $m_e=2$ KeV) and the superposed prediction
for 2 mass states with the same parameters (notice the inflexion in the
spectrum where the most massive state switches off).
} \label{fig1}
\end{figure}

Under the effects of flavor mixing, the answer depends crucially
on how flavor states conflict with Lorentz invariance. If there is
a preferred frame in nature (and Eq.~(\ref{enpref}) is valid) then
\be K_{max}=Q- m_{e}=Q- (m_1 \cos^2\theta  + m_2 \sin^2
\theta) \ee since the minimal neutrino energy is given by
Eq.~(\ref{restmass}). The spectrum is proportional to the phase
volume factor $E p E_{e}p_{e}$, so that \be
\label{specpref} \frac {dN}{dK}= C E p (Q-K)\sqrt
{(Q-K)^2f_e^2(Q-K)-M_{e}^2} \ee where $E=m+K$ and
$p=\sqrt {E^2-m^2}$ are the electron's energy and momentum.
Here $C$ is a constant and we have
neglected the Coulomb interaction between the final particles.
We have illustrated this possibility in Fig.~\ref{fig1}
(for clarity we have chosen  wildly unrealistic parameters).

\begin{figure}[tbp]
\psfig{file=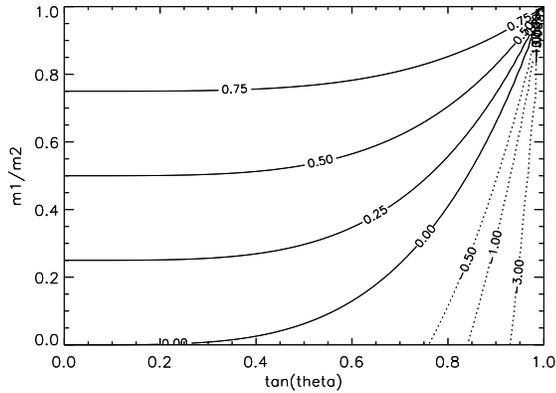,width=8cm} \caption{Contours of the parameter
$M^2$, upon which $\beta$ decay depends ($M^2$ in units of $m_2^2$).
The likelihood of any
decay experiment should follow these lines in $\{\tan(\theta),m_1/m_2\}$
space.} \label{param}
\end{figure}

However if we reject the existence of preferred frames, the
statement of energy conservation must be Eq.~(\ref{ennonli}).
Then, if $M_e^2\ge 0$, the endpoint of $\beta$ decay is  $
K_{max}=Q-M_e$, since the minimum of $E_ef_e(E_e)$ is $M_e$, the
kinematic invariant defined by Eq.~(\ref{lorentz}). The $\beta$
spectrum is now proportional to the phase volume factor $E p
E_{e}f_e(E_{e}) p_{e}$, so that: \be \frac {dN}{dK}=C
E p (Q-K)\sqrt {(Q-K)^2-M_{e}^2} \label{spec} \ee in
contradiction with Eq.~(\ref{specpref}). The only measurable
parameter is now $M^2$, and in Fig.~\ref{param} we plotted the
contours in $\{\tan(\theta),m_1/m_2\}$ space along which the
likelihood is aligned. This is to be contrasted with oscillation
experiments, which are sensitive to $\theta$ and $\Delta m^2=m^2_2-m_1^2$.
We have also plotted in Fig.\ref{fig1} predictions for $M^2_2=0,\pm1$
KeV. We see that the tail of the spectrum is distinctly different from
the case where there is a preferred frame.

The case $M_e^2<0$ merits special consideration. It does not imply
faster than light propagation, or the need for a preferred frame.
Upon closer inspection, however, we find that
$E_{e}f_e(E_{e})$ is not bounded from below. Furthermore,
imposing the condition $E_{e}f_e(E_{e})\ge 0$ (in the same
way that we have imposed $k>k_{min}$ before) violates Lorentz
invariance. Thus we obtain an unstable theory (with
$K_{max}=\infty$) unless we are prepared to accept preferred
frames (or a maximal boost parameter for any given frame).

Thus, at the level of interactions, the case $M_e^2<0$ {\it has} to
violate the principle of relativity. The minimum of $Ef(E)$ is zero, so
that the endpoint of $\beta$ decay is $K_{max}=Q$. Eq.~(\ref{spec})
is still valid, and in this case one observes an excess (rather
than a deficiency) of events near the endpoint, as compared to the
zero mass case. This seems currently to be favored by
observations~\cite{troi,mainz}.

The limiting case $M_e=0$, is on the contrary  consistent with
the principle of relativity.
Such a neutrino does not behave like a massless
particle: it travels slower than light, and has a rest frame
(unlike the case $M_e^2<0$). One can impose the condition $E_\nu
f_\nu(E_\nu)\ge 0$ without violating Lorentz invariance. So the
end point of $\beta$ decay is $K_{max}=Q$, and Eq.~(\ref{spec}) is
valid, so that the zero mass, no-mixing case is perfectly mimicked
as far as $\beta$ decay is concerned.

To conclude, all these predictions are significantly different
from those obtained by giving primacy to the mass (rather
than flavour) eigenstates~\cite{katrin}. Then the $\beta$ spectra is
\be \label{specmass} \frac {dN}{dK}= C E p E_e
\sum_i |U_{ei}|^2 \sqrt{E_e^2-m_{i}^2}\Theta (E_e-m_i) \ee
(where $E_e=Q-K$ and $U_{ei}=(\cos\theta,\sin\theta)$).
The end point is at $K=Q-m_1$ and the spectrum has an inflexion at $K=Q-m_2$.
We have also plotted this possibility in Fig.~\ref{fig1}.

In summary, we have investigated how flavor states cannot satisfy
standard dispersion relations, and studied the implications for
the principle of relativity. We found that although we may
introduce a preferred frame to describe these states, this is not
necessary as long as we are prepared to consider non-linear
realizations of the Lorentz group. The only exception is the
region of parameter space defined by Eq.~(\ref{tachcond}), where,
contrary to all appearances, one needs a preferred frame to
enforce stability.

Having laid down all possibilities, we then computed the spectrum
and endpoint of $\beta$ decay. We found a distinct prediction in each
case, a matter of great interest given prospective experimental
improvements~\cite{katrin}.
Thus the unusual interplay between Lorentz
invariance and neutrino flavor mixing is an issue to be
decided by  experimentalists.

{\bf Acknowledgements} We thank K. Baskerville
for help in connection with this paper.
M.~B. thanks the ESF network COSLAB and  EPSRC for support.

\end{document}